\documentstyle[prl,aps,epsf]{revtex} \def\narrowtext{} \tighten \twocolumn
\input epsf.sty
\begin{document}
\draft
\narrowtext
\title{Hot Spots on the Fermi Surface of Bi2212: Stripes versus Superstructure}
\author{J. Mesot,$^{1,2}$ M. R. Norman,$^1$
        H. Ding,$^{1,2}$ and J. C. Campuzano$^{1,2}$}
\address{
 (1) Materials Science Division, Argonne National Laboratory,
     Argonne, IL 60439 \\
 (2) Department of Physics, University of Illinois at Chicago,
     Chicago, IL 60607 \\}
\maketitle

In a recent paper Saini {\it et al.}\cite{SAINI1} have reported evidence 
for a pseudogap around $(\pi,0)$ at room 
temperature in the optimally doped superconductor Bi2212. This result is 
in contradiction with previous ARPES measurements\cite{NAT96}. 
Furthermore they observed at certain points on the Fermi surface hot 
spots of high spectral intensity which they relate to the existence of 
stripes in the CuO planes. They also claim\cite{SAINI1,SAINI2}
to have identified a new electronic band along $\Gamma-M_1$ whose one 
dimensional character provides further evidence for stripes.
We demonstrate in this Comment
that all the measured features can be simply understood by
correctly considering the superstructure (umklapp) and shadow
bands which occur in Bi2212. 

\begin{figure}
\epsfxsize=2.7in
\epsfbox{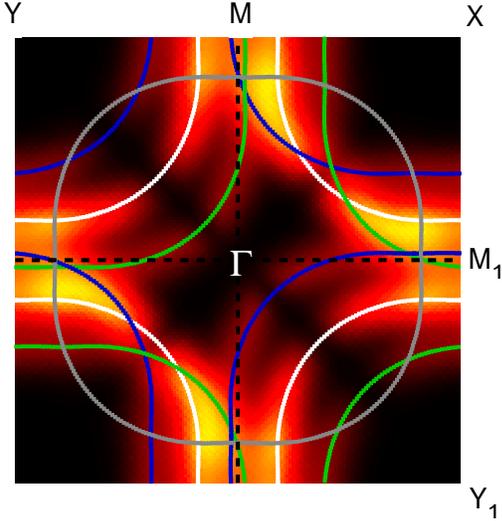}
\vspace{0.3cm}
\caption{
Plot of the calculated intensity around the Fermi energy.  Note the strong
resemblance to Fig.~1 of Ref.~1.}
\label{fig1}
\end{figure}

In Fig.~1 we plot the main Fermi surface (white) as determined from a tight
binding fit to ARPES spectra\cite{NORM95}, the $Q=\pm(0.21\pi,0.21\pi)$
umklapps (blue,green) and the
$G=(\pi,\pi)$ shadow band (gray).
The intensity plot in Fig.~1 is obtained using the following relations\cite{FT}
\begin{equation}
I_k = P_k + \alpha (P_{k+Q} + P_{k-Q}) + \beta P_{k+G}
\end{equation} 
\begin{equation}
P_X = e^{-\frac{1}{2}(\frac{X-k_F}{\eta})^2}
(|\cos(\phi)\cos(2\theta)|+|\sin(\phi)\sin(2\theta)|)
\end{equation}
where $\phi$ is the angle between $k$ and $X$ and $\theta$ the angle between
$X$ and $\Gamma-M$.
The angular term simulates the polarization dependence of the measured 
intensity and is the simplest form consistent with the dipole selection rules.
The smearing of the data due to the finite 
resolution is simulated by the parameter $\eta=0.04\pi$ in the 
exponential. The factors $\alpha=0.4$ and $\beta=0.1$ represent the relative
weights of the umklapp and shadow bands, respectively,
chosen to be consistent with those discussed in earlier work\cite{AEBI,DING96}.
Although our calculation uses a very simplistic 
matrix element, it contains the necessary ingredients to 
understand, on a semi-quantitative basis, the measured spectra.

The effect of the selection rules is well observed along $\Gamma-Y$
where no intensity is detected. 
There is spectral weight along $\Gamma-X$
due to the umklapp bands.
The hot spots are a consequence of the umklapp bands
and therefore are not an indication of stripes or a pseudogap.
In addition, the correct way to infer a
pseudogap is to compare the leading edge of the ARPES spectrum to a
Fermi function. Such ARPES measurements (including Fig.~2 of Ref.~1)
do not reveal the presence of a 
pseudogap above 100 K in optimally doped Bi2212\cite{NAT96}.

We now turn to the problem of the one-dimensional band\cite{SAINI1,SAINI2}.
As obvious from Fig.~1, the crossing of the umklapp band and
the $\Gamma-M$ axis at $(0.4\pi,0)$ is very sensitive to the sample alignment.
We note that the $M$ and $M_1$ points may appear to be
inequivalent if the rotation axis
and the sample's normal are not colinear. In that case and 
starting from the $\Gamma-X$ direction, a +45$^{\circ}$ ($M$) rotation is not
equivalent anymore to a -45$^{\circ}$ ($M_1$) rotation. 
We argue that this is what is responsible for the inequivalence seen in
Ref.~\onlinecite{SAINI1}, and that the one-dimensional band is actually the 
umklapp band.

This work was supported by the Swiss National Science Foundation, 
the U. S. Dept. of Energy,
BES, under Contract W-31-109-ENG-38, the National 
Science Foundation DMR 9624048, and
DMR 91-20000 through the Science and Technology Center for
Superconductivity.


\begin{references}
\bibitem{FT}
Eq.~2 reflects the geometry of Ref.~1 where the photon polarization
is parallel to $k$, not $X$.
\end{references}
\end{document}